\documentstyle[12pt,aasms4]{article}

\begin{document}
\title{ON A RAPID LITHIUM ENRICHMENT AND DEPLETION OF K GIANT STARS}
\author{R. de la Reza\altaffilmark {1}, N.A. Drake\altaffilmark {1,2},  
L. da Silva\altaffilmark {1}, C. A. O. Torres\altaffilmark {3} \& 
E. L. Martin\altaffilmark {4}}
\affil{$^{1}$Observat\'orio Nacional/CNPq, Departamento de Astronomia, Rua General 
 Bruce, 586, 20921-400 Rio de Janeiro, Brazil } 
\affil{$^{2}$Astronomical Institute, St. Petersburg University, St. Petersburg, 
198904 Russia}
\affil{$^{3}$Laborat\'orio Nacional de Astrof\' \i sica-CNPq, 
    MG, Brazil}
\affil{$^{4}$Instituto de Astrof\' \i sica de Canarias, E-38200 
      La Laguna, Tenerife, Spain}
\authoremail{delareza@on.br}

\begin{abstract}
A model scenario has recently been introduced to explain the presence of
very strong Li lines in the spectra of some low mass K giant stars (de la
Reza et al. 1996). In this scenario {\it all} ordinary, Li poor, K giants become
Li rich during a short time ($\sim10^5$ yr) when compared to the red giant phase 
of $5\cdot 10^7$ yr. In this ``Li period'', a large part  of the stars are
associated with an expanding thin circumstellar shell supposedly triggered by an 
abrupt internal mixing mechanism resulting in a surface new $^7$Li  enrichment. 
This letter presents near 40 Li rich K giants known up to now. 
The distribution of these Li rich giants, along with other 41 observed K 
giants that have  
shell, but are not Li rich, in a color-color {\it IRAS} diagram confirms this 
scenario, indicating, also as a new result,  that a rapid Li depletion 
takes place on a time scale of  between $\sim10^3$    
and  10$^5$ yr. This model explains the problem of the presence 
of K giants
with far infrared excesses presented by Zuckerman et al. (1995). Other present and
future tests of this scenario are briefly discussed.
\end{abstract}


\section{Introduction}

Since the discovery in the eighties of some K giant stars showing strong
lines of Li in their spectra, the main tendency has been to consider these
stars  as peculiar. This was believed  because, following classical 
first dredge-up theory (Iben 1967), all K giants would have their surface Li
depleted due to convective mixing with internal material devoid of 
this element, or  through direct Li destruction by inward transport. 
In fact, this Li depletion has 
 actually been observed for a large part of the K giants and in several 
cases, as in the  stars with masses lower than $2.5M_\odot$, 
Li depletions are larger than those indicated by theory, suggesting that an 
additional mixing mechanism could be present (Gilroy 1989). Some hypotheses 
have been formulated to try to explain the apparently anomalous presence of Li 
in the Li rich giants.  
These are: a) maintenance during the giant phase of a pre-mainsequence 
large Li abundance. b) engulfing orbiting planets and/or
brown dwarfs. c) $^{7}$Li production by a nearby nova companion. d) internal
production of fresh $^{7}$Li.

The maintenance hypothesis is very improbable due first, to mainsequence 
depletion and second, to the first dredge-up action. Also, the 
measured isotope ratios of $^{12}$C/$^{13}$C, which indicate the degree of
mixing, appear  not to be correlated with  strong Li lines; these 
lines must then be the result of recent Li production 
(da Silva, de la Reza \& Barbuy 1995).
In the engulfing as in the  maintenance 
mechanisms, there is no new Li production; Li contained in the planets
or in the brown dwarfs (Rebolo et al. 1996) is introduced in the  newly formed
giant. This mechanism could gain  new impetus due to the recent 
discoveries of extrasolar ``51 Peg'' type hot planets (Mayor \& Queloz 
1995). Nevertheless, we know neither if the Li contained in these planets
is sufficient to contaminate a star, nor what the proportions of these kinds 
of planets are when compared to the more distant ``cool'' planets also 
recently discovered. 
These more external planets are not efficient sources for Li in the engulfing process
because, due to stellar mass loss, these planets will escape  being
swallowed (Sackmann, Boothroyd, \& Kraemer 1993). Concerning Novae, even
if these objects can eventually produce nearly 15\% of the Galactic Li
(Hernanz et al. 1996), there is no observational support for the presence
of white dwarfs as hot companions to the Li K giants. This is 
based on IUE observations (de la Reza \& da Silva 1995) and on the
absence of companions indicated by the lack of  radial velocities variations
in Li K giants  at the 1 km/s level (de Medeiros, Melo, \& Mayor 1996).
Recently detected variations at the 20 m/s level in the K giant $\beta$  Ophiucus 
appear to be due rather to non-radial pulsations (Hatzes \& Cochran 1996).

We remain then with the internal production mechanism and the main
purpose of this letter consists of presenting observations supporting 
the idea  that Li K giants are not peculiar K giants, but normal giants 
going through a short Li rich period, presented for the first time in de la Reza,
Drake, \& da Silva (1996), hereafter called Paper I.

\section{The Scenario}

In Paper I, a scenario  was proposed to explain the existence of Li K giant
stars. This model was constructed based on  the discovery that almost
all Li K giants are optical counterparts of {\it IRAS} sources indicating the
presence of dusty circumstellar shells (CS).
Near half of the known Li K giants indicated in this letter were discovered 
as a subproduct of the search for new T Tauri stars,   
Pico dos Dias Survey (PDS)  made in Brazil (Gregorio-Hetem et al. 1992; 
Torres et al. 1995; de la Reza et al. 1997). A survey for new Li K giants 
in {\it IRAS} color boxes other than that corresponding to T Tauri stars has 
been initiated by Gregorio-Hetem, Castilho, \& Barbuy (1993). 
First, the presence of a CS region was believed to be, apart from Li, 
the main difference between Li K and normal K giants. However, Fekel et al. 
(1996) showed that some K giants could possess CS without showing a strong 
Li feature. Also, Zuckerman et al. (1995) presented a list of nearly 90 stars, 
the large part of which were K giants, with {\it IRAS} fluxes ratios compatible  
with the presence of CS regions. No explanation for the presence of large 
far infrared excesses in these oxygen type giants was found by these last authors.
In this paper we report on observations of 27 stars from this list in 
order to insert them into our scenario.

The main characteristics of the scenario are the following: 1) all
normal K giants (at least single field stars with masses between 
approximately 1.0
to $2.5M_\odot$) become Li rich during a short time of
$\lesssim 10^5$ yr compared  to the  red giant phase duration 
($5 \cdot 10^7$ yr). 
2) An abrupt mixing mechanism producing a rapid 
surface injection of material with fresh internal 
$^7$Be, the only new
formed element at this stage via $^3$He$+^4$He and  
rapidly transformed to $^7$Li, 
 produces the formation of a CS of gas and dust. 3) When this 
sudden mass loss stops, the CS detaches and is ejected into 
the interstellar medium.
The best values for the CS, adjusted to observations (see Paper I), give
an expansion velocity of 2 km/s and an equivalent  mass loss of 
$2-5 \cdot10^{-8}M_\odot$yr$^{-1}$ which is a hundred times the normal mass 
loss of ordinary
K giants. The complete ejection of the CS up to the stage when it is 
no longer detectable lasts at least $8\cdot 10^4$ yr.
4) The fresh $^7$Li that has not been ejected into the interstellar medium
remains in the stars' photospheres. This surface Li will be  depleted later
in a time less or equal to the total CS ejection time, depending on the stellar
parameters. This depletion mechanism is probably related inversely to the
additional mixing mechanism introduced to produce the Li enrichment.

\section{Observations}

The results presented in this paper were obtained from
spectroscopic observations  using the following telescopes: 
1) The CTIO 4.0 m in Cerro Tololo - Chile 
with an echelle spectrograph with 0.08\AA/pixel (April - May, 1996);
2) The 3.5 m of the Calar Alto Observatory 
(Almer\' \i a, Spain)  with the TWIN spectrograph with 0.88\AA/pixel 
(July, 1996); and 
3) The 2.5 m Isaac Newton  (La Palma, Spain) with the IDS spectrograph 
with 0.85\AA/pixel (November, 1996). 
The first main 
program stars consisted of a group of faint Li K and Li poor
K giants discovered in the PDS. These PDS observations were restricted to
coude spectra of a small spectral interval between H$_\alpha$ and the Li$\,$I
resonance line. The main interest in  obtaining echelle spectra was to perform
a detailed analysis  in order to derive the main stellar parameters. The
second program stars consisted of relatively bright K giants belonging to the list
of Zuckerman et al. (1995). Bearing in mind  that these stars have
{\it IRAS} colors indicating the presence of  CS regions, these 
objects were potential candidates for new Li K giants. The 
observed results presented in this letter are limited to the 
presence, or lack thereof,  of strong Li lines. A next paper will be 
devoted to the determinations of the masses and metallicities of these
objects.

\section{Discussion of the Results and Tests of the Model}

The most complete list of observed Li K giants known up to now is 
presented in Table 1, where K giants are considered  between the included 
spectral types limits of G8 and M0.
In this Table numbers with an asterisk indicate the Li K giants which are 
represented in Fig.$\;$1 with their respective labels.
From the observational point of view, we considered as Li K giant 
stars those presenting the resonance Li$\;$I line at 6708\AA$\;$ with intensities 
comparable to or higher than the neighbor Ca$\;$I at 6718\AA. 
When the Li abundance is known, ``Li K giants'' are stars with Li abundances larger
than $\log \epsilon({\rm Li})=1.2$ (where $\log \epsilon({\rm H})=12.00$). 
K giants presenting no strong Li lines and having far infrared excesses 
belong to the list of
Zuckerman et al. (1995) (with HD numbers) and to the PDS list (with {\it IRAS}
numbers).
Apart from the bright and already known Li K giants
discovered by several authors 
resulting in 21 objects,
we have added 20 new Li K giants discovered in the PDS (Gregorio-Hetem et
al. 1992; de la Reza et al. 1997) at  fainter visual magnitudes. These 
groups, bright ($m_V$ of $3^m - 8^m$) and faint  ($9^m - 14^m$), are located in different 
places in a color-color diagram with {\it IRAS} fluxes at 12, 25 and 60 $\mu$m 
(see Fig. 1). The explanation of this separation in apparent magnitudes appears
naturally in the  model  of Paper I. In that
paper is considered the existence of three regions in the diagram of
[60~$-$~25] vs [25~$-$~12] labeled I, II and III (see Fig.$\,$1).
In region I are the bright, commonly called,  normal K giants with {\it IRAS} colors of 
photospheric origin showing the absence of CS regions. It is in this 
region that almost all the Li poor giants are found. Region 
II corresponds to the color box used in the PDS to discover new T Tauri
stars. It is in this region that the new faint recently discovered  
 Li K giants are placed . In region III are found all the
previously known visual bright Li K giants. Some of the curves resulting
from the  model of Paper I are presented in Fig. 2. They represent
the evolutionary paths of the ejections of the CS regions. Beginning in 
I, they go to II and then  III before returning to I. This loop takes
at least  $8 \cdot 10^4$ yr. The stars remain for a long time in region I
during the red giant phase. 

When stars become
$^7$Li enriched in a rapid episode, a CS is formed. When the mass loss
(and the associated Li enrichment) stop, the CS detaches from the star 
ejecting its matter into the interstellar medium. In this way $^7$Li surface 
enrichment and subsequent depletion are  time adjusted with the expansion 
of the CS. Any expansion time can then be used to measure the $^7$Li 
depletion during the loops shown in Fig.$\;$2.

 There are no observed stars between regions I and II due to the very
short corresponding evolutionary times (of the order of hundreds of years).
 In region II only faint giants are observed. Even if the
corresponding  times are longer in this part of the diagram (more than one
thousand years), they are too short  to observe objects in  part II among
the nearby stars. We must then survey a larger region consisting of more distant 
and faint giants. In
region III, the crossing times are nearly ten thousand years, and these longer times
give the possibility of observing some Li K giants among  the 
bright and nearby stars.

In Paper I it was  mentioned that Li K giants contain a large majority of CS
stars. This gives us  the idea that normally Li poor K
giants have no CS regions and that the Li depletion  times were of the
order of $80\,000$ yr, or somewhat larger. The possibility that some
K giants having a CS region without high Li could exist, was pointed out 
by Fekel et al. (1996). They suggested that Li depletion times could be
smaller than $10^5$ yr,  or even that CS could be formed without Li 
enrichment. The most important result of this letter is to show that the
first conclusion is the most probable. This can be seen very clearly in
Figs. 1 and 2. There is a group  formed of Li K giants only in the lower part
of region II, having the largest [25~$-$~12] and lowest [60~$-$~25] values. 
Those stars would be the most recently formed  Li K giants! During the PDS
we found some faint K giants without high Li, however less numerous than
the Li K giants. Nevertheless, these Li poor CS K giants are all in the upper
part of region II. These stars would then be the first Li depleted giants.
All but one of the 
giants belonging to the list of Zuckerman et al. (1995) are in the  
left part of the diagram (region III and in the intersection of III and I).

In observations of 27  stars of the Zuckerman et al. list, we didn't 
discover any new Li K giants, only   Li poor K giants. These stars are then  surface
depleted Li K giants. After all, the time scales over which  depletion 
can occur are longer in region III  than 
in region II. The only star of the Zuckerman et al. (1995) list belonging to
the lower part of region II was HD$\,19745$ and this star had already been 
discovered to be very Li
rich in the PDS (Gregorio-Hetem et al. 1992, see also de la Reza \&
da Silva 1995). In conclusion, with the exception of a homogeneous group
of pure ``young'' Li rich K giants in the lower part of region II,
the Li rich and Li poor giants are mixed indicating that different Li
depletion times occur. The observed positions in Fig. 2 give 
depletion times between $\sim 10^3$ and  $10^5$ yr. Some Li K giants, such as 
HD$\,$39853 or HD$\,$787, are placed in region I showing no CS. 
Those stars are considered to have lost their CS but have not yet depleted 
their Li. The Li
K giant  HD$\,$95799 detected by Luck (1994)  is the only Li K which is not
an {\it IRAS} source. The explanation is that this star is also in region I, like
HD$\,$39853 and HD$\,$787, but is not detected by {\it IRAS} because it  is more
distant  than 
these last ones. In fact, the visual magnitude of HD$\,$95799 is $m_V=8.01$,
whereas those of HD$\,$39853 and HD$\,$787 are respectively 5.66 and 5.25.

Li enrichment and depletion in K giants appears then to be the 
explanation to the problem posed by Zuckerman et al. (1995) with respect to the 
existence of oxygen giants with strong far infrared excesses. Is stellar mass the factor which
determines a shorter or longer depletion time? Determinations of the main
stellar parameters of these CS stars, with and without high Li, could
give an answer to this question. Other important parameters, such as a strong
differential rotation (Fekel et al. 1996) or metallicity could, however, 
have a significant  part in this rapid enrichment-depletion process.

Some questions have no answer yet. Can stars become Li rich
several times during the giant phase? This, in principle, depends on the
quantity of available $^3$He,  which acts as a fuel for the production and
enrichment of $^7$Be, subsequently transformed into $^7$Li. Direct 
observational evidence of this could be shown by the  eventual detection of double
detached CS. Another question refers to the internal nature of these stars.
Are they first ascent red giant or clump giants? We do not propose 
an answer to this yet, due to  uncertainties related to the positions of
these stars in the HR diagram. As can be seen in Paper I, these stars are
somewhat grouped in the HR diagram. Future accurate determinations of the
stellar parameters will elucidate this question. Other considerations are 
important; this ``Li phenomenon'' has been observed only for single
field giants spread over all Galactic latitudes. No clear 
identification of a Li K giant belonging to a cluster has yet been made.
This will be important to determine the ``Li age'' for these stars.
From the theoretical point of view, considerable advance has been made 
on the stellar internal $^7$Li production and surface enrichment in low
mass  K giants $(1 - 2.5M_\odot)$ by means of an internal 
circulation mechanism (Sackmann \& Boothroyd 1997). This one relates the
base of the convective layer to a hot region producing the 
$^7$Be. Larger Li abundances can be obtained depending upon certain values
of the circulation mechanism, 
as the mixing speed, the depth of mixing, the star's metallicity, and 
possibly the star's mass. For example, large Li abundances
as $\log \epsilon(^7$Li$)=4.2$ were obtained for a $1M_\odot$ star of 
metallicity $Z=0.0001$ with a rapid mixing to greater depths.

Perhaps one of the most interesting points of the scenario of Paper I
is that it can be tested by several means. One has  already been made
by Fekel et al. (1996) by means of the determination of the real nature of 
the star HDE$\,$233517. This object was considered by Skinner et al. 
(1995) and Miroshnichenko, Bergner \& Kuratov (1996) to be a nearby K 
dwarf  star of the ``Vega'' or ``$\beta$ 
Pictoris'' type. Fekel at al. (1996) showed that this star is in reality a 
distant K giant with a large mass loss (object number 15 in
Fig.$\,$1). According to  its position in the color-color diagram this 
star should be probably Li 
rich and, according to  its corresponding mass loss and age of its shell, its CS
should be of a  specific size. Both points have been confirmed by 
observations (Fekel et al. 1996). Other general tests can 
be made such as detection
of the sizes and velocities of the CS, presence of CS detachment and/or
double shells. Also, observations of the presence of $^9$Be, $^{10,11}$B and $^6$Li
will give an insight into the rate   of the $^7$Li enrichment process
(Sackmann \& Boothroyd 1997). A Non LTE Li abundance determination of
several prototype stars in different regions of the color-color diagram is in
progress. These determinations will help us to quantify their Galactic
Li enrichment contribution.

\acknowledgments
We would like to thank V. Smith and D.L. Lambert for helpful discussions, and 
I.-J. Sackmann and A.I. Boothroyd  for sending us their important  recent 
research results.


\begin{figure}
\caption{Distribution of {\it IRAS} sources corresponding to K giants  contained 
  in three regions labeled I, II, and III marked by solid lines. The regions marked by
  broken lines are those defined by  van der Veen \& Habing (1988).   
  Labeled filled  symbols correspond to {\it Li strong K giants}: squares 
  (three fluxes are of good quality), 
  triangles  (one flux is only an upper limit) and hexagons (two fluxes are 
  only  upper limits). Corresponding open symbols
  represent {\it Li weak K giants}. Crosses represent K giant stars of the 
  Zuckerman et al. (1995) list with CS regions and which 
  have not yet been observed in the spectral Li region.}
\end{figure}

\begin{figure}
\caption{The same points as in  Fig.$\;$1 are presented here together with  evolution  curves
  of CS (see Paper I). The four curves calculated for CS expansion velocity of 2 km/s,
  stars temperatures and radii equal to 4000$\,$K and  $20R_\odot$ respectively, 
  correspond to mass losses between $10^{-9}$ and $10^{-6}M_\odot$/yr.
  The blackbody curve (straight dashed line) and time steps are also indicated.}
\end{figure}


\begin{table}
\caption{List of K giants observed in the Lithium spectral region.} 
\begin{tabular}{rl@{\qquad}rl@{\qquad}rl}

 1$^\ast$ & {\rm HD$\,$787}   & 29$^\ast$ & {\rm HD$\,$112127}                & 57$\;\,$ & {\rm HD$\,$162298} \\
 2$^\ast$ & {\rm HD$\,$9746}  & 30$^\ast$ & {\rm IRAS$\,13313\!\!-\!\!5838$}  & 58$\;\,$ & {\rm IRAS$\,17576\!\!-\!\!1845$} \\
 3$^\ast$ & {\rm CPD-55395}   & 31$\;\,$ & {\rm HD$\,$118344}                & 59$^\ast$ & {\rm IRAS$\,17578\!\!-\!\!1700$} \\
 4$^\ast$ & {\rm HD$\,$19745} & 32$\;\,$ & {\rm HD$\,$119853}                & 60$^\ast$ & {\rm IRAS$\,17582\!\!-\!\!2619$} \\
 5$^\ast$ & {\rm IRAS$\,03520\!\!-\!\!3857$} & 33$^\ast$   & {\rm HD$\,$120602} & 61$\;\,$ & {\rm IRAS$\,17590\!\!-\!\!2412$} \\ 
 6$^\ast$ & {\rm HD$\,$30238} & 34$^\ast$ & {\rm PDS$\,68$}                   & 62$^\ast$ & {\rm IRAS$\,17596\!\!-\!\!3952$} \\  
 7$^\ast$ & {\rm HD$\,$30834} & 35$^\ast$ & {\rm HD$\,$121710}                & 63$\;\,$ & {\rm HD$\,$164712} \\
 8$^\ast$ & {\rm HD$\,$31993} & 36$\;\,$ & {\rm HD$\,$125618}                & 64$^\ast$ & {\rm IRAS$\,18334\!\!-\!\!0631$} \\
 9$^\ast$ & {\rm HD$\,$39853} & 37$\;\,$ & {\rm HD$\,$128309}                & 65$\;\,$ & {\rm IRAS$\,18397\!\!-\!\!0400$} \\
10$\;\,$ & {\rm IRAS$\,06365\!\!+\!\!0223$} & 38$\;\,$ & {\rm HD$\,$129955} & 66$\;\,$ & {\rm IRAS$\,18559\!\!+\!\!0140$} \\
11$^\ast$ & {\rm IRAS$\,07227\!\!-\!\!1320$} & 39$\;\,$ & {\rm HD$\,$131530} & 67$^\ast$ & {\rm HD$\,176588$} \\
12$^\ast$ & {\rm IRAS$\,07456\!\!-\!\!4722$} & 40$\;\,$ & {\rm IRAS$\,14198\!\!-\!\!6115$} & 68$^\ast$ & {\rm IRAS$\,19012\!\!-\!\!0742$} \\
13$^\ast$ & {\rm IRAS$\,07577\!\!-\!\!2806$} & 41$\;\,$ & {\rm IRAS$\,14257\!\!-\!\!6023$} & 69$\;\,$ & {\rm HD$\,177366$} \\
14$^\ast$ & {\rm HD$\,$65750} & 42$^\ast$ & {\rm IRAS$\,16086\!\!-\!\!5255$}  &  70$^\ast$ & {\rm IRAS$\,19083\!\!+\!\!0119$} \\
15$^\ast$ & {\rm HDE$\,$233517} & 43$^\ast$ & {\rm IRAS$\,16128\!\!-\!\!5109$} & 71$\;\,$ & {\rm HD$\,181154$} \\
16$\;\,$ & {\rm HD$\,$76066}   & 44$\;\,$ & {\rm HD$\,$146834}              & 72$\;\,$ & {\rm IRAS$\,19210\!\!+\!\!1715$} \\
17$\;\,$ & {\rm HD$\,$82227}   & 45$^\ast$ & {\rm HD$\,$146850}              & 73$\;\,$ & {\rm HD$\,183202$} \\
18$\;\,$ & {\rm HD$\,$82421}   & 46$^\ast$ & {\rm HD$\,$148293}              & 74$^\ast$ & {\rm PDS$\,$100} \\
19$\;\,$ & {\rm IRAS$\,09553\!\!-\!\!5621$} & 47$\;\,$ & {\rm IRAS$\,16227\!\!-\!\!4839$} & 75$\;\,$ & {\rm HD$\,$187114} \\
20$\;\,$ & {\rm HD$\,$92253}   & 48$\;\,$ & {\rm IRAS$\,16252\!\!-\!\!5440$} & 76$\;\,$ & {\rm HD$\,$190299} \\
21$\;\,$ & {\rm HD$\,$95799}   & 49$^\ast$ & {\rm IRAS$\,16514\!\!-\!\!4625$} & 77$^\ast$ & {\rm HD$\,$194317}\\
22$^\ast$ & {\rm IRAS$\,11044\!\!-\!\!6127$} & 50$\;\,$ & {\rm HD$\,$153135}  & 78$^\ast$ & {\rm HD$\,$203251} \\
23$\;\,$ & {\rm HD$\,$96996} & 51$\;\,$ & {\rm HD$\,$156061}                 & 79$\;\,$ & {\rm HD$\,$204540} \\
24$\;\,$ & {\rm HD$\,$97472} & 52$\;\,$ & {\rm HD$\,$156115}                 & 80$\;\,$ & {\rm HD$\,$218527} \\
25$^\ast$ & {\rm IRAS$\,12236\!\!-\!\!6302$} & 53$\;\,$ & {\rm IRAS$\,17102\!\!-\!\!3813$} & 81$^\ast$ & {\rm HD$\,$219025} \\
26$^\ast$ & {\rm HD$\,$108471}  & 54$\;\,$ & {\rm IRAS$\,17120\!\!-\!\!4106$}               & 82$\;\,$ & {\rm HD$\,$221776} \\
27$^\ast$ & {\rm IRAS$\,12327\!\!-\!\!6523$} & 55$^\ast$ & {\rm IRAS$\,17211\!\!-\!\!3458$} \\
28$\;\,$ & {\rm HD$\,$111830}  & 56$\;\,$ & {\rm IRAS$\,17442\!\!-\!\!2441$} \\   

\end{tabular}
\end{table}

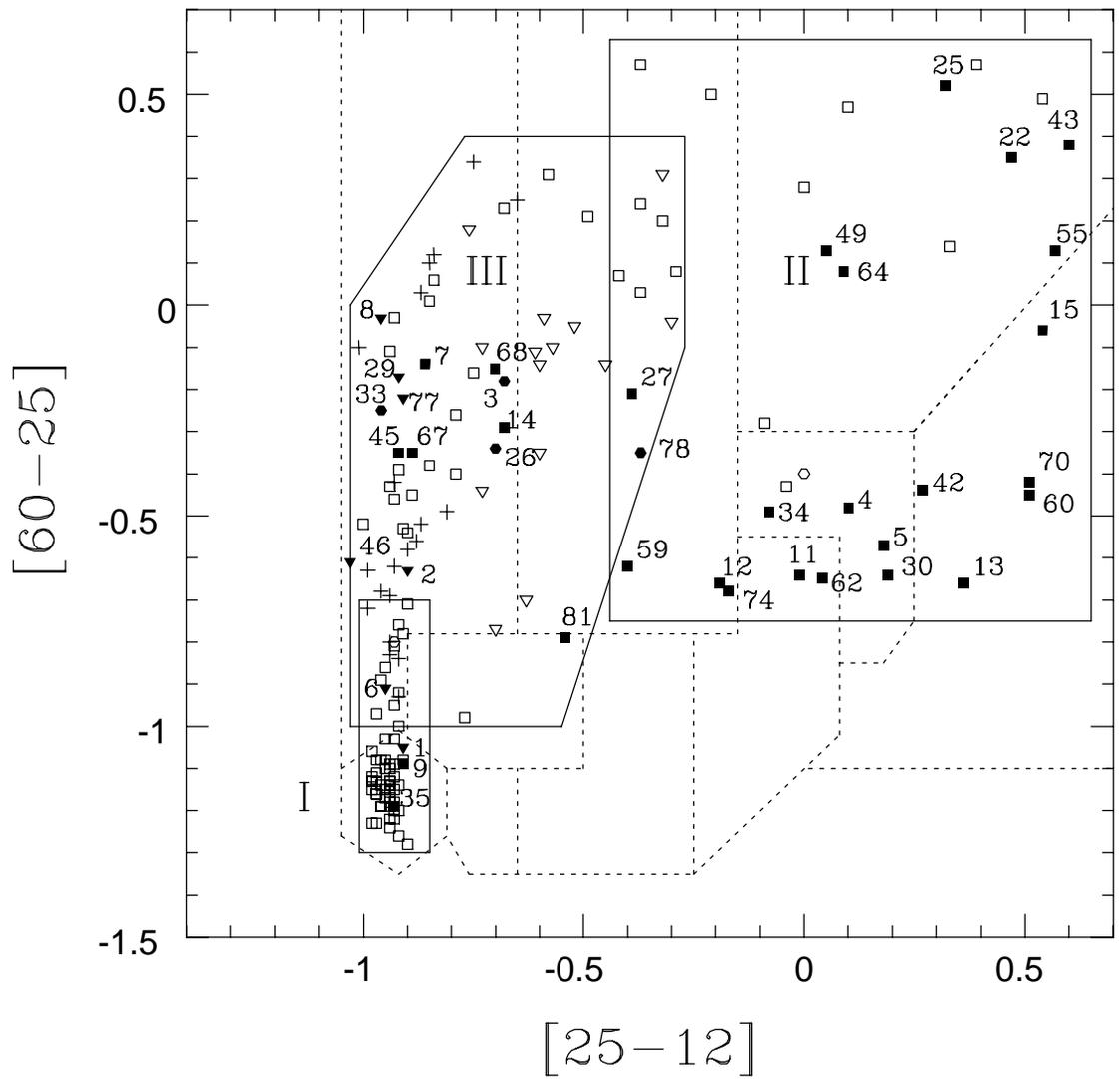
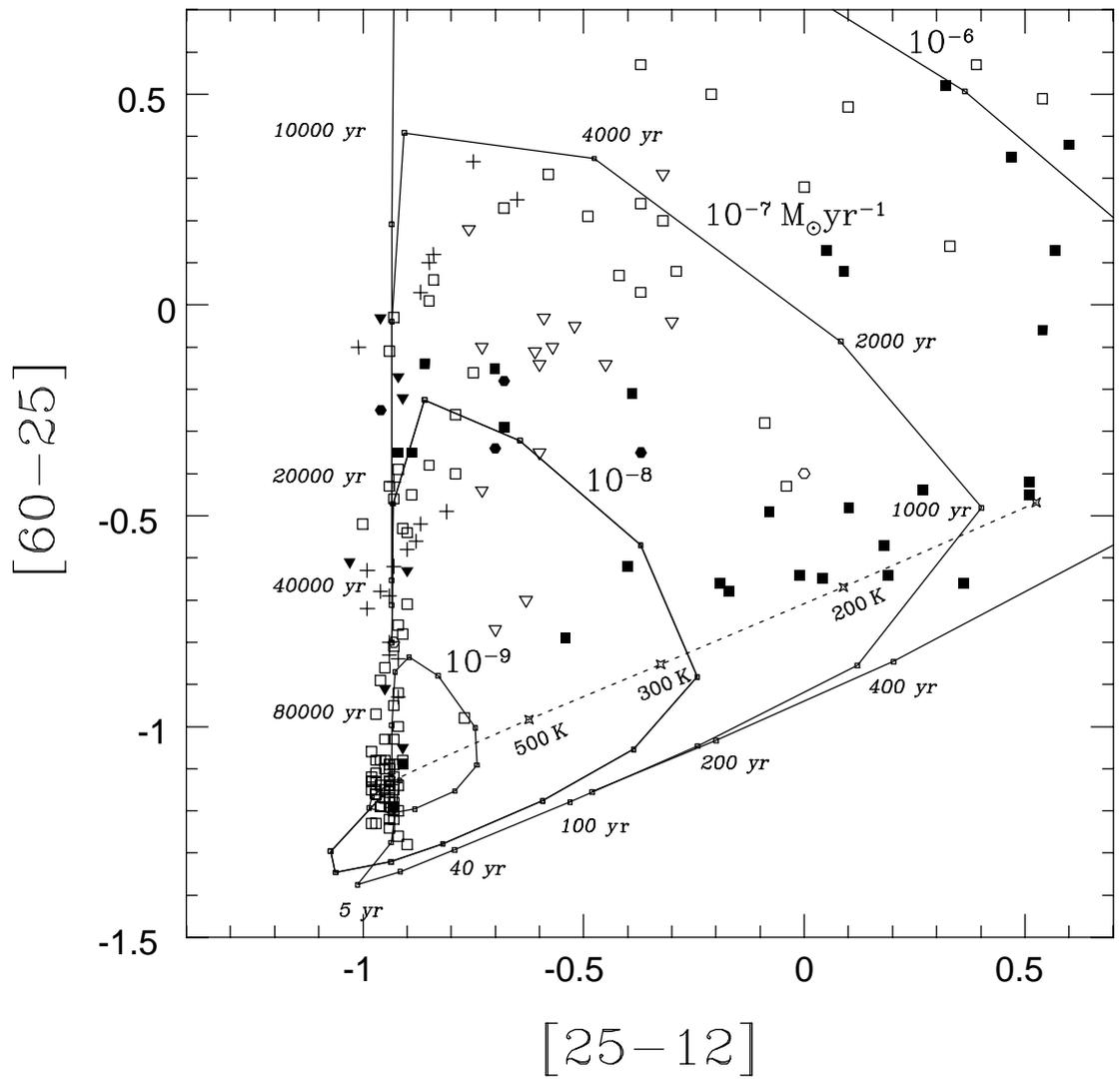
\end{document}